\documentclass{article}
\usepackage{spconf,amsmath,graphicx}
\usepackage{amssymb}

\title{SHAPE AWARE AUTOMATIC REGION-OF-INTEREST SUBDIVISIONS}
 \name{Timothy L. Kline}
 \address{Department of Radiology,\\
Mayo Clinic,
Rochester MN, 55905 USA \\
    }
\begin{document}
%
\maketitle
\begin{abstract}
In a wide variety of fields, analysis of images involves defining a region and measuring its inherent properties. Such measurements include a region's surface area, curvature, volume, average gray and/or color scale, and so on. Furthermore, the subsequent subdivision of these regions is sometimes performed. These subdivisions are then used to measure local information, at even finer scales. However, simple griding or manual editing methods are typically used to subdivide a region into smaller units. The resulting subdivisions can therefore either not relate well to the actual shape or property of the region being studied (i.e., gridding methods), or be time consuming and based on user subjectivity (i.e., manual methods). The method discussed in this work extracts subdivisional units based on a region's general shape information. We present the results of applying our method to the medical image analysis of nested regions-of-interest of myocardial wall, where the subdivisions are used to study temporal and/or spatial heterogeneity of myocardial perfusion. This method is of particular interest for creating subdivision regions-of-interest (SROIs) when no variable intensity or other criteria within a region need be used to separate a particular region into subunits.  \end{abstract}
\begin{keywords}
Computed Tomography, Fast Marching Method, Nested Region-of-interest, Shape-based Analysis, Subdivision
\end{keywords}
\section{Introduction}
\label{sec:intro}

Image analysis frequently involves the localization and/or segmentation of a particular region located within an image. These regions-of-interest (ROIs) are used to measure such properties as an object's surface area, curvature, volume, average gray and/or color scale, etc. Methods for automatically segmenting such regions as the heart, kidney, liver, and lungs, have been researched exhaustively\cite{Soler,Zheng}. However, a common step in image analysis also includes the subdivision of a region into even smaller components\cite{Ritman,Stanley,Kim,King}. The methods for subdividing a region-of-interest, or subdivision ROIs (SROIs), have previously necessitated user interaction. These SROIs are then used to measure local information at even finer scales. 

The method developed in this paper automatically subdivides a region into equal area sections based on local shape information. The method has the potential to greatly speed up the analysis process, and also remove errors, compared to simple griding or manual editing methods. The method, as presented, could easily be extended to a wide variety of image analysis problems where the characterization of regional subunits is used. Therefore, some potential extensions to the current method are also discussed.

\begin{figure}[!t]
\centering
\includegraphics[width=8.5cm]{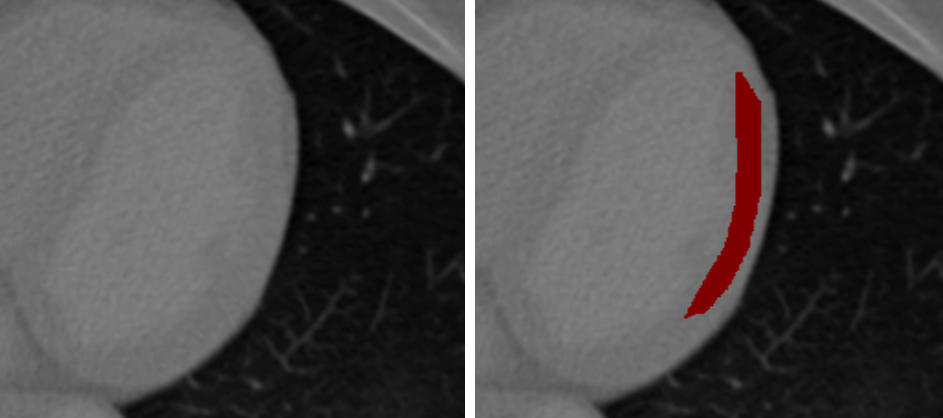}
\caption{Left panel: cross-sectional slice of a CT image of human myocardium. Right panel: user-defined region composed of myocardial wall.}
\label{fig_sim}
\end{figure}

\begin{figure}[!t]
\centering
\includegraphics[width=8.5cm]{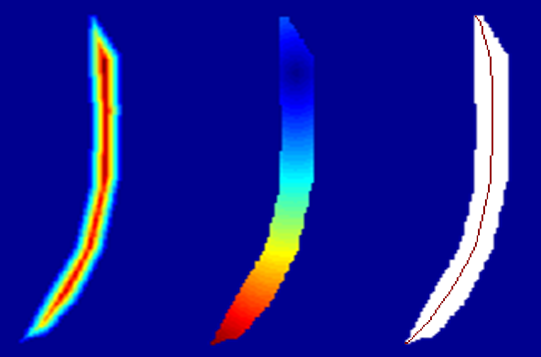}
\caption{Depicts various steps during the automated centerline extraction method. Beginning with the binary mask the distance map is computed (left panel). Utilizing the fast marching algorithm (middle panel), the ends of the object are automatically determined. A gradient descent method is then used to extract the objects centerline (right panel). This centerline serves as a guide for extracting sub-divisions of the region of interest.}
\label{fig_sim}
\end{figure}

\section{Methods}

\subsection{Region-of-Interest}
Our method to delineate subregions uses a binary mask as input. This mask is of equal size as the image under analysis. The mask assigns voxels a value of one if contained by the region, and a value of zero otherwise. Shown in Fig. 1 is a CT cross-sectional image of a human myocardium, here used as the test image to illustrate the steps of our method. Also, the right panel of Fig. 1 displays a user created region-of-interest in a left-ventricule-free portion of the myocardial wall. Such an ROI would subsequently be subdivided in order to study temporospatial heterogeneity of myocardial perfusion distribution\cite{Ritman,King}.  

\subsection{Subdivision}

\subsubsection{Region Centerline}
\noindent The binary data set of the ROI was used as the input for extracting the region's centerline, and is similar to a method previously presented\cite{klineABME}. Using locations along the centerline, perpendicular cuts are made to then distinguish individual subunits of the region. The first step involves computing the distance map\cite{maurer} of the ROI, as shown in the upper left corner of Fig. 2. The distance map assigns each voxel of a region a value equal to the Euclidean distance between that voxel and the nearest non-region voxel. Therefore, the innermost voxels have the highest value. 

The fast marching algorithm \cite{28} was then utilized to locate one of the object's ends (Fig. 2 upper right), and then a second time (Fig. 2 lower left) to aid in computing a discrete, shortest path curve that defines the region's centerline (Fig. 2 lower right). Briefly, the fast marching algorithm begins at a source (in this case a single voxel) and simulates a wave emanating in all directions. The propagating wave is controlled by a weighting map (i.e., the distance map) which tells the algorithm how fast the wave travels through each voxel. As the wave passes over each voxel, ``how long'' it took the wave to get to that voxel is recorded, termed the ``arrival time''. Therefore, the arrival time is a function of both the distance a particular voxel is from the source of the wave front, as well as the weighting function that the wave front experiences along the way. 

In $\mathbb{R}^2$ we begin with a potential function $V(x)$ that weights a region's voxels according to their distance from the region's border. The weighted geodesic distance between two points $x_0,x_1 \in \mathbb{R}^2$ is given by\cite{peyre}

\begin{equation}
U(x_0,x_1) = \min\limits_\gamma\left(\int_0^1{\|\gamma'(t)\|V(\gamma(t))dt}\right),
\end{equation}

\noindent where $\gamma$ is the geodesic curve with $\gamma(0) = x_0$, $\gamma(1) = x_1$, and $\|$ represents the Euclidean norm (i.e., the vector length). In the case that $V(x) = 1 \forall x$ then the integral in Eq. 3 is simply the length of $\gamma$ and $U$ is the Euclidean (straight-line) distance between $x_0$ and $x_1$. Solving the Eikonal equation 
\noindent 

\begin{equation} \label{GrindEQ__1_} 
\left\|\nabla U(x) \right\|= V(x) ,                                                   
\end{equation} 

\noindent where \textit{U} is the arrival time function, \textit{V }the potential function or the ``speed'' of the front progression, and $\nabla$ the gradient operator, the weighted geodesic distance map from a starting point voxel (point automatically chosen as discussed below) to all other voxels inside the ROI was computed. This equation is solved by the fast marching method by computing \textit{U} at location \textit{(x,y)} using the arrival time value of its neighbors. An easily implementable ``upwind'' approach\cite{rouy} was used.

Shortest paths can then be extracted from any voxel to the wave front's source voxel (thus finding the geodesic curve $\gamma$ from $x_0$ to $x_1$)\cite{88}. A discrete gradient descent method was utilized to compute this path\cite{sedgewick}.

Fig. 2 depicts snapshots along the centerline defining algorithm. The distance map of the region is first computed (Fig. 2 upper left) and the maximum value found is used as the starting point for the first fast marching wave (Fig. 2 upper right). The maximum value found from this fast marching wave is then used to propagate the second and final fast marching wave (which is weighted by the distance map to the sixth power so that the wave travels fastest in the region's center), as shown in the lower left panel of Fig. 2. Finally a discrete gradient descent is performed starting from the maximum voxel in the image created by the second fast marching wave. This shortest path defines the object's centerline as shown in the lower right panel of Fig. 2 overlaid on the original binary mask.

\subsubsection{Divisions}

To divide the ROI into subdivisions, cross-sectional areas, or line-profiles were calculated at each centerline voxel. In our current example we illustrate the division of a single ROI into 16 subregions. First, 15 evenly spaced voxels are labeled in the ordered list of centerline voxels. For each of the 15 voxels, the normal vector, or direction of the centerline, is computed from the two adjacent voxels. The adjacent voxels of \textit{($x_{i}$,$y_{i}$)} are labelled  \textit{($x_{i-1}$,$y_{i-1}$)} and \textit{($x_{i+1}$,$y_{i+1}$)}. The normal vector is then calculated as

\begin{equation} \label{GrindEQ__6_} 
N=(x_{i+1} - x_{i+1} )\hat{x}+(y_{i+1} - y_{i-1} )\hat{y}.
\end{equation} 

\noindent where $\hat{x}$, and $\hat{y}$ are unit vectors in the two cardinal directions. The next step is to determine voxels in the volume that lie perpendicular to the normal vector that are connected to the centerline voxel. These are voxels (at location ($x$,$y$)) satisfying

\begin{equation} \label{GrindEQ__7_} 
(x_{i+1} -x_{i-1} )(x-x_{i-1} )\hat{x}+(y_{i+1} -y_{i-1} )(y-y_{i-1} )\hat{y} = 0.     \end{equation}

Three example cross-sections, or line profiles cutting through the region, are shown in the first three panels of Fig. 3. Separating the object by these cuts, the individual regions can be delineated by a connected component analysis. To label the individual regions, the algorithm iterated through each cut, and added the removed region to the existing image. The result was a mask with regions labeled from 1 to 16, thereby defining the SROIs (right panel of Fig. 3). 

\begin{figure}[!t]
\centering
\includegraphics[width=8.5cm]{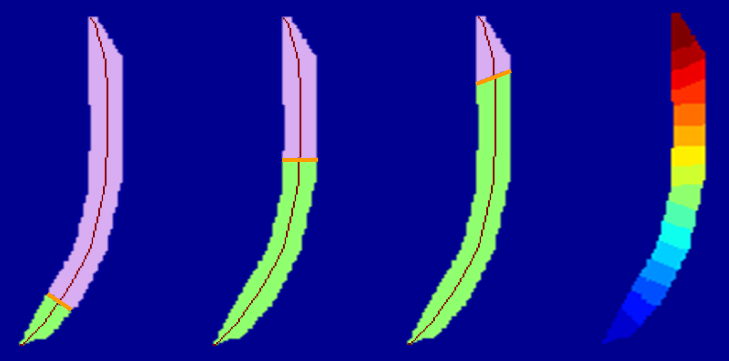}
\caption{Example of the perpendicular cutting process. The normal vector is first computed by solving Eq. 3. Next, the voxels that satisfy Eq. 4 are flagged. These are shown as the orange voxels cutting through the region perpendicular to the region's centerline. The region is seperated by this perpendicular cut into two regions. The disconnected region is added to the binary mask image at each iteration resulting in the output SROIs being represented by values between 1 and 16. The result is a new image mask in which the single ROI has been divided into 16 sub-divisions that follow the curvature of the original object.}
\label{fig_sim}
\end{figure}

\subsubsection{Area Correction}
In the case of certain types of analysis, it may also be desirable to keep the area (or volume) of each subregion the same. To create equal area regions an additional step was implemented. First, the largest region was considered (call $R_l$). Next, the two neighboring regions' areas were computed and the smaller of the two was chosen (call $R_s$). All voxels of $R_l$ bordering with $R_s$ were changed to be labeled as part of $R_s$. This process was repeated (until no new $R_l$'s were chosen) in order to perform a rough area correction that retains the shape of the individual regions (only changing the width). Finally the small differences between areas were corrected by exchanging bordering voxels from larger regions with those of neighboring smaller regions. Since the area of the ROI will most likely not be a multiple of the number of SROIs to be generated, a final automation step removed a few outer voxels of the last region (starting from those with a maximum value in the second fast marching wave image) so that each of the regions have equal areas. 

\begin{figure}[htb]

\begin{minipage}[b]{0.48\linewidth}
  \centering
  \centerline{\includegraphics[width=4.0cm]{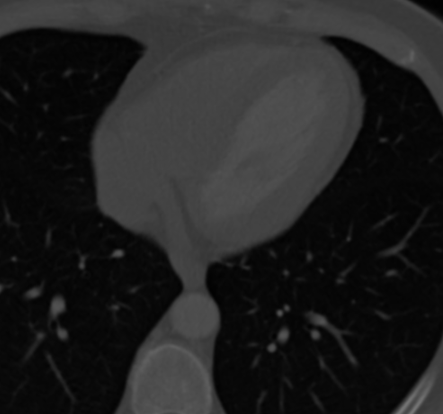}}
  \centerline{(a)}\medskip
\end{minipage}
\begin{minipage}[b]{.48\linewidth}
  \centering
  \centerline{\includegraphics[width=4.0cm]{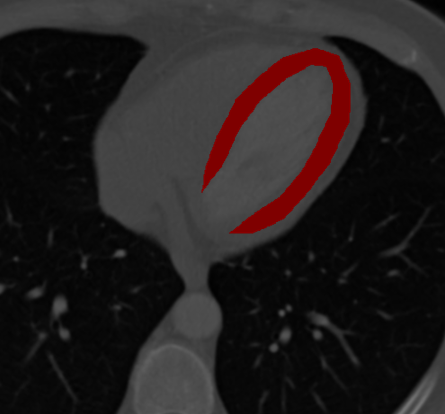}}
  \centerline{(b)}\medskip
\end{minipage}
\hfill
\begin{minipage}[b]{1.0\linewidth}
  \centering
  \centerline{\includegraphics[width=8.5cm]{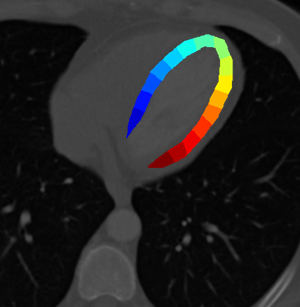}}
  \centerline{(c)}\medskip
\end{minipage}
\caption{Example of SROIs generated for an annular shaped ROI. (a) CT gray scale image of human myocardium. (b) User-defined ROI. (c) Result of the automatic region-of-interest subdivision algorithm. Here the region has been divided into 16 equal area regions which follow the curvature of the original object.}
\label{fig:res}
\end{figure}

\section{Results}
The application of this method to a second image where the ROI is annulus shaped is shown in Fig. 4. Notice that the subdivisions retain a shape that follows the general shape of the ROI, and each SROI is of equal area. Similar shaped ROIs could also be used to study other physiological properties such as local perfusion defects and characterize such properties as regions of hyper- or hypo- perfusion in vessel walls \cite{Cury}. The SROIs defined manually by this method are visually similar to those found in applied studies \cite{Dong}. 

\section{Discussion}
The developed method discussed in this work is straight-forward to implement and robust at extracting SROIs that pertain to the general shape of the input ROI. The particular method, as presented, involved creating a specified number of same area SROIs. The method could easily be extended to a variety of problems. For instance, it may be of interest to segment regions based on local curvature (i.e., regions may not be constrained to have equal areas). Therefore, an analysis of the change in the computed normal vector along the centerline could be used to locate points of high curvature, and extract SROIs based on these points. Similarly, creating SROIs based on the change in cross-sectional area or diameter along the centerline could be another means of creating SROIs that are delineated based on their local shape change information. The method is an effective means of creating SROI regions when no variable intensity within a region can be, or need be, used to delineate subunits of a particular region. 

\bibliographystyle{IEEEbib}

\end{document}